# Stochastic Calculus and the Black-Scholes-Merton Model: A Simplified Approach

Kuo-Ping Chang*

* Jinhe Center for Economic Research, Xi'an Jiaotong University, Xianning West Road, 28#, Xi'an, Shaanxi Province, 710049, China. E-mail: kpchang@mx.nthu.edu.tw.

Stochastic Calculus and the Black-Scholes-Merton Model: A Simplified Approach

## ABSTRACT

In the continuous-time finance literature, it is claimed that the expected rate of return of the underlying asset does not affect the option pricing model. This paper shows that under no arbitrage, i.e., under the Arbitrage (Gordan) theorem, the use of different underlying asset price processes in the Black-Scholes-Merton partial differential equation and the Black-Scholes-Merton option pricing formula requires that the risk-free interest rate be a linear function of the underlying asset's expected rate of return (alpha) and its return variance, or that (as in the literature) the risk-free interest rate equal the underlying asset's alpha. This result indicates possible mispricing in options markets and the potential for arbitrage opportunities.

## 1. Introduction

The seminal works of Black and Scholes (1973) and Merton (1973) have inspired many researches on pricing and hedging different financial contracts. The literature argues that the option pricing formula does not contain underlying asset's expected rate of return (e.g., Hull, 2022, Ross, 1993, Shreve, 2004, and Steele, 2001). This paper has shown that under no arbitrage, i.e., under the Arbitrage (Gordan) theorem, the use of different underlying asset price processes in the Black-Scholes-Merton partial differential equation and the Black-Scholes-Merton option pricing formula require that the risk-free interest rate be a linear function of the underlying asset's expected rate of return (alpha) and its return variance, or that (as in the literature) the risk-free interest rate equal the underlying asset's alpha.

The remainder of this paper is organized as follows. Section 2 introduces the Arbitrage Theorem in the framework of the binomial option pricing model. No arbitrage in the Black-Scholes-Merton model is shown in Section 3. Concluding remarks appear in Section 4.

## 2. Derivation of the Binomial Option Pricing Model

As shown in Figure 1, at $t = 0$, we have a stock with current price $S(0) = S_0 = \$100$ and a European call $c(0) = c_0$ of three-month with exercise price $K = \$110$ and a European put option $p(0) = p_0$ of three-month with exercise price $K = \$110$. Suppose that after three months, at $t = T$, the stock price will either go up to $S_0 u = \$120$ or go down to $S_0 d = \$80$ and $r = 0.03$ is the continuously compounded risk-free rate of interest.

$$c_u = Max[S_0 u - K, 0] = Max[120 - 110, 0]$$

$$p_u = Max[K - S_0 u, 0] = Max[110 - 120, 0]$$

$$S_0 u = 120$$

$$S_0 = 100 \qquad e^r = e^{0.03} \qquad c_0 = ? \qquad p_0 = ?$$

$$S_0 d = 80$$

$$c_d = Max[S_0 d - K, 0] = Max[80 - 110, 0]$$

$$p_d = Max[K - S_0 d, 0] = Max[110 - 80, 0]$$

Figure 1. A binomial option pricing model

Then from the Arbitrage (Gordan) theory we have:[1]

$$\begin{bmatrix} e^r1 - e^r1 & e^r1 - e^r1 \\ S_0u - e^rS_0 & S_0d - e^rS_0 \\ c_u - e^rc_0 & c_d - e^rc_0 \\ p_u - e^rp_0 & p_d - e^rp_0 \end{bmatrix} \begin{bmatrix} \pi \\ 1-\pi \end{bmatrix} = \begin{bmatrix} e^{0.03} - e^{0.03} & e^{0.03} - e^{0.03} \\ 120 - e^{0.03}100 & 80 - e^{0.03}100 \\ 10 - e^{0.03}c_0 & 0 - e^{0.03}c_0 \\ 0 - e^{0.03}p_0 & 30 - e^{0.03}p_0 \end{bmatrix} \begin{bmatrix} \pi \\ 1-\pi \end{bmatrix} = \begin{bmatrix} 0 \\ 0 \\ 0 \\ 0 \end{bmatrix}, \tag{1}$$

$$\boldsymbol{A^T} \qquad \boldsymbol{p} \qquad\qquad = \boldsymbol{0}$$

hence, $\pi = 0.5761$, $c_0 = 5.5911$, and $p_0 = 12.34$.

Eq. (1) can be rewritten as:[2]

$$\begin{cases} \text{Money Market:} & e^r \cdot \pi + e^r \cdot (1-\pi) = e^{0.03} \cdot \pi + e^{0.03} \cdot (1-\pi) = e^{0.03} \cdot 1 \\ \text{Stock:} & S_0u \cdot \pi + S_0d \cdot (1-\pi) = 120\pi + 80(1-\pi) = e^{0.03}100 = E_{\boldsymbol{P}}[S(T)|S_0] = e^r \cdot S_0 \\ \text{Call:} & c_u \cdot \pi + c_d \cdot (1-\pi) = 10 \cdot \pi + 0 \cdot (1-\pi) = e^{0.03}5.5911 = E_{\boldsymbol{P}}[c(T)|c_0] = e^r \cdot c_0 \\ \text{Put:} & p_u \cdot \pi + p_d \cdot (1-\pi) = 0 \cdot \pi + 30 \cdot (1-\pi) = e^{0.03}12.34 = E_{\boldsymbol{P}}[p(T)|p_0] = e^r \cdot p_0 \end{cases} \tag{2}$$

---

[1] As shown in Chang (2015), the Arbitrage (Gordan) theory is:

Let $\boldsymbol{A}$ be an $m \times n$ matrix. Then, exactly one of the following systems has a solution:

System 1: $\boldsymbol{Ax} > \boldsymbol{0}$ for some $\boldsymbol{x} \in \boldsymbol{R^n}$

System 2: $\boldsymbol{A^T p} = \boldsymbol{0}$ for some $\boldsymbol{p} \in \boldsymbol{R^m}$, $\boldsymbol{p} \geq \boldsymbol{0}$, $\boldsymbol{e^T p} = 1$ where $\boldsymbol{e} = \begin{bmatrix} 1 \\ 1 \\ \cdot \\ \cdot \\ \cdot \\ 1 \end{bmatrix}$.

If System 2 holds (i.e., no arbitrage) and the matrix $\boldsymbol{A}$ has rank $m-1$ (i.e., a complete market), the probability measure $\boldsymbol{p}$ will be unique. (Suppose that $\boldsymbol{A^T p} = \boldsymbol{0}$ and $\boldsymbol{p} = (\pi_1, \ldots, \pi_m)^{\boldsymbol{T}}$ is a non-zero vector. Then the rank of $\boldsymbol{A^T}$, $R(\boldsymbol{A^T})$, is less than $m$. Unique solution for $(\pi_1, \ldots, \pi_m)$ and $\sum_{i=1}^{m} \pi_i = 1$ imply $R(\boldsymbol{A^T}) = m - 1$.) Note that System 2 is a martingale.

[2] For a more extensive discussion of the properties of the binomial option pricing model (e.g., option Greeks), see Chang (2023).

where $\pi = \frac{e^r - d}{u-d}$ , $1-\pi = \frac{u-e^r}{u-d}$ , and $\boldsymbol{p} = (\pi, 1-\pi)$ is the probability measure governing the stochastic process.

Eq. (1) shows that it is a complete market. In a complete market, the complete set of possible bets on future states of the world can be constructed with existing assets without friction. For example, at $t = 0$ we can form a portfolio $X(0)$ which contains $\Delta$ shares of the stock and money $B$ in a bank with the continuously compounded risk-free rate of interest $r = 0.03$ to replicate the call's future payoff at $t = T$ (i.e., a hedging strategy):[3]

$$\begin{cases} \Delta \cdot S_0 u + e^r B = c_u = 10 \\ \Delta \cdot S_0 d + e^r B = c_d = 0 \end{cases} , \tag{3}$$

where $\Delta = \frac{c_u - c_d}{S_0 u - S_0 d} = 0.25$. Thus, $B = e^{-r}[c_u - \Delta \cdot S_0 u] = e^{-r}\left[\frac{u \cdot c_d - d \cdot c_u}{u-d}\right] = -19.4089$, and with no arbitrage,

$$c_0 = X(0) = \Delta \cdot S_0 + B = e^{-r}\left[\frac{e^r - d}{u-d} \cdot c_u + \frac{u-e^r}{u-d} \cdot c_d\right] = e^{-r}[\pi c_u + (1-\pi)c_d]$$

$$= e^{-r}[\pi(\Delta \cdot S_0 u + e^r B) + (1-\pi)(\Delta \cdot S_0 d + e^r B)] = 5.5911.$$

As in eq. (2), we have:

$$(\Delta \cdot S_0 u + e^r B)(\pi) + (\Delta \cdot S_0 d + e^r B)(1-\pi) = E_{\boldsymbol{P}}[X(T)|X(0)] = e^r c_0 = e^{0.03} 5.5911 = e^r X(0). \tag{4}$$

---

[3] Alternatively, at $t = 0$, we can buy $\Delta$ shares of the stock and sell one call to construct a portfolio which gives a certain future payoff at $t = T$, and $\begin{cases} 120(\Delta) - 110 = 80(\Delta) - 0 \Rightarrow \Delta = 0.25 \\ \frac{80(0.25)}{e^{0.03}} = 100(0.25) - c_0 \Rightarrow c_0 = 5.5911 \end{cases}$. That is, $\Delta \cdot S_0 u - c_u = \Delta \cdot S_0 d - c_d \Rightarrow \Delta = \frac{c_u - c_d}{S_0 u - S_0 d}$, and hence, $[S_0 u \cdot \frac{c_u - c_d}{S_0(u-d)} - c_u]/e^r = S_0 \cdot \frac{c_u - c_d}{S_0(u-d)} - c_0 \Rightarrow c_0 = [\pi c_u + (1-\pi)c_d]/(1+r)$, where $\pi = \frac{e^r - d}{u-d}, 1-\pi = \frac{u-e^r}{u-d}$.

## 3. Derivation of the Black-Scholes-Merton Model

Case 1: Geometric Brownian motion process: $\{S(t) = S_0 \cdot e^{\sigma \cdot W(t)+\alpha t}, T \geq t \geq 0\}$.

Denote the stock price at time $0 \leq t_i \leq T$ as $S(t_i)$ where $S(0) \equiv S_0$ and $0 = t_0 < t_1 <\ldots< t_{n-1} < t_n = T$. Let

$$S(T) = \frac{S(T)}{S(t_{n-1})} \cdot \frac{S(t_{n-1})}{S(t_{n-2})} \cdots \frac{S(t_2)}{S(t_1)} \cdot \frac{S(t_1)}{S(0)} \cdot S(0),$$

and $y(t_n) = S(T)/S(t_{n-1})$, …, $y(t_{n-i}) = S(t_{n-i})/S(t_{n-i-1})$, $n > i \geq 1$,

hence,

$$S(T) = y(t_n) \cdot y(t_{n-1}) \cdot \ldots \cdot y(t_1) \cdot S(0) \quad \text{or} \quad lnS(T) = \textstyle\sum_{i=1}^{n} ln\big(y(t_i)\big) + ln\big(S(0)\big).$$

Suppose that $ln\,(y(t_i))$, $i \geq 1$, are independent and identically distributed. Then, assume $y(t_i)$, $i \geq 1$, are log-normal (or with large *n*, and being suitably normalized approximately be Brownian motion with a drift), $\sum_{i=1}^{n} ln\,(y(t_i)) \equiv Y(T)$ is normally distributed, i.e., $\sigma \cdot W(T) + \alpha T \equiv Y(T) \sim N(\alpha T, \sigma^2 T)$, where $W(t), t \geq 0$ is a Brownian motion and $W(t) \sim N(0,t)$. Also, $S(T) = e^{\sum_{i=1}^{n} ln(y(t_i)) + ln\,S(0)} = S_0 \cdot e^{Y(T)} = S_0 \cdot e^{\sigma \cdot W(T)+\alpha T}$ where $ln(S(0)) \equiv lnS_0$. Thus, we have $(Y(T) - Y(S)) \sim N(\alpha(T-S), \sigma^2(T-S))$.

We can show that $\left\{X(t) = S_0 \cdot e^{Y(t)-(\alpha t+\frac{1}{2}\sigma^2 t)}, T \geq t \geq 0\right\} = \left\{X(t) = S_0 \cdot e^{\sigma \cdot W(t)+\alpha t-(\alpha t\ \frac{1}{2}\sigma^2 t)}, T \geq t \geq s \geq 0\right\}$ is an exponential martingale:

$$E[X(t)|X(u), 0 \le u \le s] = E[S_0 \cdot e^{Y(t)-(\alpha t+\frac{1}{2}\sigma^2 t)}|X(u), 0 \le u \le s]$$

$$= S_0 \cdot e^{-(\alpha t+\frac{1}{2}\sigma^2 t)} \cdot E[e^{Y(t)}|X(u), 0 \le u \le s]$$

$$= S_0 \cdot e^{-(\alpha t+\frac{1}{2}\sigma^2 t)} \cdot E[e^{Y(t)-Y(s)+Y(s)}|X(u), 0 \le u \le s]$$

$$= S_0 \cdot e^{Y(s)-(\alpha t+\frac{1}{2}\sigma^2 t)} \cdot E[e^{Y(t)-Y(s)}|X(u), 0 \le u \le s]$$

$$= S_0 \cdot e^{Y(s)-\left(\alpha t+\frac{1}{2}\sigma^2 t\right)} \cdot E\left[e^{Y(t)-Y(s)}\right]$$

$$= S_0 \cdot e^{Y(s)-(\alpha t+\frac{1}{2}\sigma^2 t)} \cdot e^{\alpha(t-s)+\frac{1}{2}\sigma^2(t-s)}$$

$$= S_0 \cdot e^{Y(s)-(\alpha s+\frac{1}{2}\sigma^2 s)}$$

$$= X(s).$$

Let $s = 0$, the above equation becomes:

$$E[S_0 \cdot e^{Y(t)-(\alpha t+\frac{1}{2}\sigma^2 t)}|X(0)] = E[S(t) \cdot e^{-(\alpha t+\frac{1}{2}\sigma^2 t)}|X(0)] = S(0) = S_0.$$

Multiply both sides of the above equation by $e^{\alpha t+\frac{1}{2}\sigma^2 t}$. Then, as shown in eq. (2), i.e., with no arbitrage (the last equal sign), we have:

$$e^{\alpha t \ \ \frac{1}{2}\sigma^2 t} \cdot E[S(t) \cdot e^{-(\alpha t+\frac{1}{2}\sigma^2 t)}|X(0)] = E[S(t)|X(0)] = e^{\alpha t+\frac{1}{2}\sigma^2 t} S_0 = e^{rt} S_0. \tag{5}$$

Thus, $\{S(t) = S_0 \cdot e^{\sigma \cdot W(t)+\alpha t}, T \ge t \ge 0\}$ and no arbitrage imply:

$$\alpha + \frac{1}{2}\sigma^2 = r. \tag{6}$$

Case 2: Geometric Brownian motion process: $\{S(t) = S_0 \cdot e^{\sigma \cdot W(t)+(\alpha t - \frac{1}{2}\sigma^2 t)}, T \geq t \geq 0\}$.

Suppose that $\{S(t) = S_0 \cdot e^{\sigma \cdot W(t)+(\alpha t-\frac{1}{2}\sigma^2 t)}, T \geq t \geq 0\}$. Let $0 \cdot t + \sigma \cdot W(t) \equiv Y(t) \sim N(0, \sigma^2 t)$ where $W(t), t \geq 0$ is a Brownian motion and $W(t) \sim N(0, t)$. We can show that $\{X(t) = S_0 \cdot e^{Y(t)+\left(\alpha t-\frac{1}{2}\sigma^2 t\right)-\alpha t}, T \geq t \geq s \geq 0\}$ is an exponential martingale:

$$E[X(t)|S(u), 0 \leq u \leq s] = E[S_0 \cdot e^{Y(t)+\left(\alpha t-\frac{1}{2}\sigma^2 t\right)-\alpha t}|X(u), 0 \leq u \leq s]$$

$$= S_0 \cdot e^{(\alpha t-\frac{1}{2}\sigma^2 t)-\alpha t} \cdot E[e^{Y(t)}|X(u), 0 \leq u \leq s]$$

$$= S_0 \cdot e^{(\alpha t-\frac{1}{2}\sigma^2 t)-\alpha} \cdot E[e^{Y(t)-Y(s)+Y(s)}|X(u), 0 \leq u \leq s]$$

$$= S_0 \cdot e^{Y(s)+(\alpha t\ \ \frac{1}{2}\sigma^2 t)-\alpha t} \cdot E[e^{Y(t)-Y(s)}|X(u), 0 \leq u \leq s]$$

$$= S_0 \cdot e^{Y(s)+(\alpha t-\frac{1}{2}\sigma^2 t)-\alpha t} \cdot E\left[e^{Y(t)-Y(s)}\right]$$

$$= S_0 \cdot e^{Y(s)+(\alpha t\ \ \frac{1}{2}\sigma^2 t)-\alpha} \cdot e^{0\cdot(t-s)+\frac{1}{2}\sigma^2(t-s)}$$

$$= S_0 \cdot e^{Y(s)+\left(\alpha s-\frac{1}{2}\sigma^2 s\right)-\alpha s}$$

$$= S(s) \cdot e^{-\alpha}$$

$$= X(s).$$

Let $s = 0$, the above equation becomes:

$$E[S_0 \cdot e^{Y(t)+\left(\alpha t-\frac{1}{2}\sigma^2 t\right)-\alpha t}|X(0)] = E[S(t) \cdot e^{-\alpha t}|X(0)] = S(0) = S_0.$$

Multiply both sides of the above equation by $e^{\alpha t}$. Then, as shown in eq. (2), i.e., with no arbitrage (the last equal sign), we have:

$$e^{\alpha t} \cdot E[S(t) \cdot e^{-\alpha t}|X(0)] = E[S(t)|X(0)] = e^{\alpha t}S_0 = e^{rt}S_0. \tag{7}$$

Thus, $\{S(t) = S_0 \cdot e^{\sigma \cdot W(t)+(\alpha t-\frac{1}{2}\sigma^2 t)}, T \geq t \geq 0\}$ and no arbitrage imply:

$$\alpha = r. \tag{8}$$

Case 3: Geometric Brownian motion process: $\{S(t) = S_0 \cdot e^{\sigma \cdot W(t)-\frac{1}{2}\sigma^2 t}, T \geq t \geq 0\}$.

Suppose that $Y(t) \sim N(0, \sigma^2 t)$ and $\sigma \cdot W(t) \equiv Y(t) \sim N(0, \sigma^2 t)$ where $W(t), t \geq 0$ is a Brownian motion and $W(t) \sim N(0, t)$. Then $\{S(t) = S_0 \cdot e^{Y(t)-\frac{1}{2}\sigma^2 t}, T \geq t \geq s \geq 0\}$ is an exponential martingale:

$$\begin{aligned}
E[S(t)|S(u), 0 \leq u \leq s] &= E[S_0 \cdot e^{Y(t)-\frac{1}{2}\sigma^2 t}|S(u), 0 \leq u \leq s] \\
&= S_0 \cdot e^{-\frac{1}{2}\sigma^2 t} \cdot E[e^{Y(t)}|S(u), 0 \leq u \leq s] \\
&= S_0 \cdot e^{-\frac{1}{2}\sigma^2 t} \cdot E[e^{Y(t)-Y(s)+Y(s)}|S(u), 0 \leq u \leq s] \\
&= S_0 \cdot e^{Y(s)-\frac{1}{2}\sigma^2 t} \cdot E[e^{Y(t)-Y(s)}|S(u), 0 \leq u \leq s] \\
&= S_0 \cdot e^{Y(s)-\frac{1}{2}\sigma^2 t} \cdot E\left[e^{Y(t)-Y(s)}\right] \\
&= S_0 \cdot e^{Y(s)-\frac{1}{2}\sigma^2 t} \cdot e^{0 \cdot (t-s)+\frac{1}{2}\sigma^2 (t-s)} \\
&= S_0 \cdot e^{Y(s)-\frac{1}{2}\sigma^2 s} \\
&= S(s).
\end{aligned}$$

Let $s = 0$, the above equation becomes:

$$E[S_0 \cdot e^{Y(t)-\frac{1}{2}\sigma^2 t}|S(0)] = E[S(t)|S(0)] = S(0) = S_0.$$

As shown in eq. (2), i.e., with no arbitrage (the last equal sign), we have:

$$E[S(t)|S(u), 0 \leq u \leq s] = e^{0 \cdot t} S_0 = e^{rt} S_0. \tag{9}$$

Thus, $\{S(t) = S_0 \cdot e^{\sigma \cdot W(t) - \frac{1}{2}\sigma^2 t}, T \geq t \geq 0\}$ and no arbitrage imply:

$$r = 0. \tag{10}$$

We now adopt $\{S(t) = S_0 \cdot e^{\sigma \cdot W(t) + \alpha t}, T \geq t \geq 0\}$ in Case 1 as a stock price process.[4] Let $X(t) = \sigma \cdot W(t) + \alpha t$, and $S(t) = f(X(t))$, where $f(x) = S(0) \cdot e^x$, $f'(x) = S(0) \cdot e^x$ and $f''(x) = S(0) \cdot e^x$. We have: $dX(t) = \sigma \cdot dW(t) + \alpha \cdot dt$. By Ito's lemma,

---

[4] In the literature, $\{S(t) = S_0 \cdot e^{\sigma \cdot W(t) + (\alpha t - \frac{1}{2}\sigma^2 t)}, T \geq t \geq 0\}$ in Case 2 is used. Let $X(t) = \sigma \cdot W(t) + (\alpha t - \frac{1}{2}\sigma^2 t)$. We have: $dX(t) = \sigma \cdot dW(t) + (\alpha - \frac{1}{2}\sigma^2)dt$. By Ito's lemma,

$$dS(t) = df(X(t)) = f'(X(t)) \cdot dX(t) + \frac{1}{2} f''(X(t)) \cdot dX(t) \cdot dX(t) = S(t) \cdot dX(t) + \frac{1}{2} S(t) \cdot dX(t) \cdot dX(t)$$

$$= (\alpha - \frac{1}{2}\sigma^2) \cdot S(t) \cdot dt + \sigma \cdot S(t) \cdot dW(t) + \frac{1}{2} \cdot S(t) \cdot \sigma^2 dt = \alpha \cdot S(t) \cdot dt + \sigma \cdot S(t) \cdot dW(t).$$

$$dS(t) = df(X(t)) = f'(X(t)) \cdot dX(t) + \frac{1}{2} f''(X(t)) \cdot dX(t) \cdot dX(t)$$

$$= S(t) \cdot dX(t) + \frac{1}{2} S(t) \cdot dX(t) \cdot dX(t)$$

$$= \alpha \cdot S(t) \cdot dt + \sigma \cdot S(t) \cdot dW(t) + \frac{1}{2} \cdot S(t) \cdot \sigma^2 dt$$

$$= [\alpha \cdot S(t) + \frac{1}{2} \cdot S(t) \cdot \sigma^2] dt + \sigma \cdot S(t) \cdot dW(t). \tag{11}$$

and

$$d(e^{-rt} S(t)) = df(t, S(t))$$

$$= f_t(t, S(t)) \cdot dt + f_x(t, S(t)) \cdot dS(t) + \frac{1}{2} f_{xx}(t, S(t)) \cdot dS(t) \cdot dS(t)$$

$$= -re^{-rt} S(t) \cdot dt + e^{-rt} \cdot dS(t) + 0$$

$$= (\alpha + \frac{1}{2}\sigma^2 - r) e^{-rt} \cdot S(t) \cdot dt + \sigma \cdot e^{-rt} \cdot S(t) \cdot dW(t). \tag{12}$$

Since it is a complete market, as in the discrete time case (i.e., eq.'s (3) and (4)) we can also form a portfolio valued at $X(t)$ which contains $\Delta$ shares of the stock and $X(t) - \Delta(t)S(t)$ invested in a bank with interest rate *r* to replicate the call. Then,

$$dX(t) = \Delta(t) \cdot dS(t) + r(X(t) - \Delta(t) \cdot S(t)) dt$$

$$= \Delta(t)((\alpha \cdot S(t) + \frac{1}{2} \cdot S(t) \cdot \sigma^2) dt + \sigma \cdot S(t) \cdot dW(t)) + r(X(t) - \Delta(t) \cdot S(t)) dt$$

$$= r \cdot X(t) \cdot dt + \Delta(t) \cdot (\alpha + \frac{1}{2}\sigma^2 - r) \cdot S(t) dt + \Delta(t) \cdot \sigma \cdot S(t) \cdot dW(t) \tag{13}$$

and

$$d(e^{-rt} X(t)) = df(t, X(t))$$

$$= -re^{-rt} X(t) \cdot dt + e^{-rt} \cdot dX(t) + 0$$

$$= \Delta(t) \cdot [(\alpha + \frac{1}{2}\sigma^2 - r) e^{-rt} \cdot S(t) \cdot dt + \sigma \cdot e^{-rt} \cdot S(t) \cdot dW(t)] = \Delta(t) \cdot d(e^{-rt} S(t)). \tag{14}$$

Let the value of a European call option at time $t$ be $c(t,S(t))$. We have:

$$dc(t,S(t)) = c_t(t,S(t)) \cdot dt + c_x(t,S(t)) \cdot dS(t) + \frac{1}{2} c_{xx}(t,S(t)) \cdot dS(t) \cdot dS(t)$$

$$= [c_t(t,S(t)) + (\alpha + \frac{1}{2}\sigma^2) \cdot S(t) \cdot c_x(t,S(t)) + \frac{1}{2}\sigma^2 \cdot S^2(t) \cdot c_{xx}(t,S(t))]dt$$

$$+\sigma \cdot S(t) \cdot c_x(t,S(t)) \cdot dW(t) \qquad (15)$$

and

$$d(e^{-r}\ c(t,S(t))) = df(t,c(t,S(t)))$$

$$= f_t(t,c(t,S(t))) \cdot dt + f_x(t,c(t,S(t))) \cdot dc(t,S(t))$$

$$+\frac{1}{2} f_{xx}(t,c(t,S(t))) \cdot dc(t,S(t)) \cdot dc(t,S(t))$$

$$= e^{-rt}[-r \cdot c(t,S(t)) + c_t(t,S(t)) + (\alpha + \frac{1}{2}\sigma^2) \cdot S(t) \cdot c_x(t,S(t))$$

$$+\frac{1}{2}\sigma^2 \cdot S^2(t) \cdot c_{xx}(t,S(t))]dt + e^{-rt} \cdot \sigma \cdot S(t) \cdot c_x(t,S(t)) \cdot dW(t). \qquad (16)$$

To equate the evolutions of the portfolio value and the call option value for all $t$, i.e.,

$$X(t) = e^{rt}X(0) = e^{rt}c(0,S(0)) = c(t,S(t)) \text{ or}$$

$$e^{-rt}X(t) = X(0) = c(0,S(0)) = e^{-r}\ c(t,S(t)) \text{ for all } t \in [0,T],$$

we need

$$d(e^{-r}\ X(t)) = d(e^{-rt}c(t,S(t))) \text{ for all } t \in [0,T) \qquad (17)$$

$$\text{and} \quad X(0) = c(0,S(0)).$$

Thus, by substituting eq. (14) and eq. (16) into eq. (17) we have:

$$\Delta(t) \cdot (\alpha + \tfrac{1}{2}\sigma^2 - r) \cdot S(t) \cdot dt + \Delta(t) \cdot \sigma \cdot S(t) \cdot dW(t)$$

$$= [-r \cdot c(t, S(t)) + c_t(t, S(t)) + (\alpha + \tfrac{1}{2}\sigma^2) \cdot S(t) \cdot c_x(t, S(t))$$

$$+ \tfrac{1}{2}\sigma^2 \cdot S^2(t) \cdot c_{xx}(t, S(t))]dt + \sigma \cdot S(t) \cdot c_x(t, S(t)) \cdot dW(t).$$

For the $dW(t)$ terms: $\Delta(t) = c_x(t, S(t))$.

For the $dt$ terms:

$$\Delta(t) \cdot (\alpha + \tfrac{1}{2}\sigma^2 - r) \cdot S(t) = c_x(t, S(t)) \cdot (\alpha + \tfrac{1}{2}\sigma^2 - r) \cdot S(t)$$

$$= -r \cdot c(t, S(t)) + c_t(t, S(t)) + (\alpha + \tfrac{1}{2}\sigma^2) \cdot S(t) \cdot c_x(t, S(t)) + \tfrac{1}{2}\sigma^2 \cdot S^2(t) \cdot c_{xx}(t, S(t)). \quad (18)$$

By cancelling $[(\alpha + \frac{1}{2}\sigma^2) \cdot S(t) \cdot c_x(t, S(t)]$ or substituting $\alpha + \frac{1}{2}\sigma^2 = r$ of eq. (6) into both sides of eq. (18), we have:[5]

$$r \cdot c\big(t, S(t)\big) = c_t\big(t, S(t)\big) + r \cdot S(t) \cdot c_x\big(t, S(t)\big) + \tfrac{1}{2}\sigma^2 \cdot S^2(t) \cdot c_{xx}(t, S(t))$$

$$\text{for all } t \in [0, T),\ S(t) \geq 0. \quad (19)$$

---

[5] Note that by adopting $\{S(t) = S_0 \cdot e^{\sigma \cdot W(t) + (\alpha t \ \frac{1}{2}\sigma^2 t)}, T \geq t \geq 0\}$ and hence, $dS(t) = \alpha \cdot S(t) \cdot dt + \sigma \cdot S(t) \cdot dW(t)$ in the literature, we can use the non-arbitrage requirement $\alpha = r$ of eq. (8) to derive exactly the same eq. (19).

Eq. (19) is the Black-Scholes-Merton partial differential equation.[6]

## 4. Concluding Remarks

In the continuous-time finance literature, it is claimed that the expected rate of return of the underlying asset does not affect the option pricing model. This paper has shown that under no arbitrage, i.e., under the Arbitrage (Gordan) theorem, the use of different underlying asset price processes in the Black-Scholes-Merton partial differential equation and the Black-Scholes-Merton option pricing formula require that the risk-free interest rate be a linear function of the underlying asset's expected rate of return (alpha) and its variance of return, or that (as in the literature) the risk-free interest rate equal the underlying asset's alpha. This result indicates possible mispricing in options markets and the potential for arbitrage opportunities.

---

[6] See Appendix for using the no arbitrage requirement: $\alpha + \frac{1}{2}\sigma^2 = r$ to derive the Black-Scholes-Merton option pricing formula.

# Appendix

We assume that as in Case 1, $\{S(t) = S_0 \cdot e^{\sigma \cdot W(t)+\alpha t}, T \geq t \geq 0\}$ is a stock price process, where $\sigma W(t) + \alpha t \equiv Y(t) \sim N(\alpha t, \sigma^2 t)$. At $t = T$, the value of the call option $c(T, S(T))$ is:

$$(S(T) - K)^+ = \begin{cases} S(T) - K & \text{if } S(T) \geq K \\ 0 & \text{if } S(T) < K \end{cases}$$

As shown in Case 1, let $X(T) = S_0 \cdot e^{Y(T)-(\alpha T\ \frac{1}{2}\sigma^2 T)} = S_0 \cdot e^{\sigma \cdot W(T)+\alpha T-(\alpha T\ \frac{1}{2}\sigma^2 T)} = S(T) \cdot e^{-(\alpha T+\frac{1}{2}\sigma^2 T)}$, and with no arbitrage we have eq. (5), i.e.,

$$E_{\boldsymbol{p}}[S(T)|X(0)] = e^{rT} S_0 \text{ and hence, } E_{\boldsymbol{p}}[(S(T) - K)^+|X(0)] = e^{rT} c(0, S(0)),$$

or

$$0 = E_{\boldsymbol{p}}[(S(T) - K)^+ - e^{rT} c(0, S(0))|c(0, S(0))] = E_{\boldsymbol{p}}[(S(T) - K)^+ - e^{rT} c(0, S(0))] \qquad \text{(A1)}$$

where $\boldsymbol{p}$ is the probability measure governing the stochastic process. With $S(T) = S_0 \cdot e^{Y(T)}$ and $Y(T) \sim N(\alpha T, \sigma^2 T)$, eq. (A1) can be shown as:

$$e^{rT} c(0, S(0)) = E_{\boldsymbol{p}}[(S(T) - K)^+] = \int_{-\infty}^{+\infty} (S_0 \cdot e^y - K)^+ \cdot \frac{1}{\sqrt{2\pi\sigma^2 T}} \cdot e^{\frac{-(y-\alpha T)^2}{2\sigma^2 T}} dy.$$

Rewrite $S_0 \cdot e^y - K \geq 0$ as $y \geq ln(\frac{K}{S_0})$, we have:

$$e^{rT} c(0, S(0)) = \int_{ln(\frac{K}{S_0})}^{+\infty} (S_0 \cdot e^y - K) \cdot \frac{1}{\sqrt{2\pi\sigma^2 T}} \cdot e^{\frac{-(y-\alpha T)^2}{2\sigma^2 T}} dy.$$

Let $z = (y - \alpha T)/\sigma\sqrt{T}$. We have: $dy = \sigma\sqrt{T}dz$ and

$$e^{rT}c(0, S(0)) = S_0 \cdot e^{\alpha T} \cdot \frac{1}{\sqrt{2\pi}} \int_a^{+\infty} e^{\sigma z\sqrt{T}} \cdot e^{\frac{-z^2}{2}} dz - K \cdot \frac{1}{\sqrt{2\pi}} \int_a^{+\infty} e^{\frac{-z^2}{2}} dz \qquad \text{(A2)}$$

where $a = [ln(\frac{K}{S_0}) - \alpha T]/\sigma\sqrt{T}$, and

$$\frac{1}{\sqrt{2\pi}} \int_a^{+\infty} e^{\sigma z\sqrt{T}} \cdot e^{\frac{-z^2}{2}} dz = e^{\frac{T\sigma^2}{2}} \cdot \frac{1}{\sqrt{2\pi}} \int_a^{+\infty} e^{\frac{-(z-\sigma\sqrt{T})^2}{2}} dz$$

$$= e^{T\sigma^2/2} \cdot \text{Prob}\{N(\sigma\sqrt{T}, 1) \geq a\} = e^{T\sigma^2/2} \cdot \text{Prob}\{N(0,1) \geq a - \sigma\sqrt{T}\}$$

$$= e^{T\sigma^2/2} \cdot \text{Prob}\{N(0,1) \leq \sigma\sqrt{T} - a\} = e^{T\sigma^2/2} \cdot \varphi(\sigma\sqrt{T} - a).$$

Thus, eq. (A2) becomes:

$$e^{rT}c(0, S(0)) = S_0 \cdot e^{(\alpha+\sigma^2/2)T} \cdot \varphi(\sigma\sqrt{T} - a) - K \cdot \varphi(-a). \qquad \text{(A3)}$$

By substituting the no arbitrage requirement: $\alpha + \frac{1}{2}\sigma^2 = r$ of eq. (6) into eq. (A3), we have the Black-Scholes-Merton model:

$$c(0, S(0)) = S_0 \cdot \varphi(\sigma\sqrt{T} - a) - K \cdot e^{-r} \quad \cdot \varphi(-a) = S_0 \cdot \varphi(d_1) - K \cdot e^{-r} \quad \cdot \varphi(d_2) \qquad \text{(A4)}$$

where $d_1 = \frac{ln(\frac{S_0}{K}) + (r + \frac{\sigma^2}{2})T}{\sigma\sqrt{T}}$, $d_2 = d_1 - \sigma\sqrt{T}$.

Note that if we assume, as in the literature, that $\{S(t) = S_0 \cdot e^{\sigma \cdot W(t) + (\alpha t \;\; \frac{1}{2}\sigma^2 t)}, T \geq t \geq 0\}$ is a stock price process (where $\sigma \cdot W(t) \sim N(0, \sigma^2 t)$), then with the no arbitrage requirement: $\alpha = r$ of eq. (8), we will still get the same eq. (A4).